\documentclass[12pt]{article}
\usepackage{graphicx}%
\usepackage{amssymb}
\usepackage{wrapfig}
\usepackage{amsmath}
\usepackage{booktabs}
\usepackage{color}
\renewcommand{\thefootnote}{*\arabic{footnote}}

\makeatletter
\@addtoreset{equation}{section}
\makeatother

\begin{document}
\begin{titlepage}

\begin{flushright}
\today
\end{flushright}

\vspace{4ex}

\begin{center}

{\large \bf 
CEDM constraints on modified sfermion universality\\ and spontaneous CP violation}

\vspace{6ex}

\renewcommand{\thefootnote}{\alph{footnote}}

M. Ishiduki,
S.-G. Kim\footnote{sunggi@eken.phys.nagoya-u.ac.jp},
N. Maekawa\footnote{maekawa@eken.phys.nagoya-u.ac.jp}, and
K. Sakurai\footnote{sakurai@eken.phys.nagoya-u.ac.jp}

\vspace{4ex}
{\it Department of Physics, Nagoya University,\\ Nagoya 464-8602, Japan}\\

\end{center}

\renewcommand{\thefootnote}{\arabic{footnote}}
\setcounter{footnote}{0}
\vspace{6ex}

\begin{abstract}

We discuss the supersymmetric CP problem that arises when the sfermion soft mass universality is modified.
We place the 3rd generation SU(5) ten-plet sfermion masses in the weak scale in view of the naturalness.
The other sfermion masses are assumed to be universal and a TeV scale in order to weaken the flavor changing neutral current processes and electric dipole moment (EDM) constraints.
However this modification generically induces too large up quark chromo-EDM (CEDM) via the weak scale stop loop.
In order to suppress this CEDM, we propose certain type of flavor structure where the parameters of the up-(s)quark sector are real whereas those of the down-(s)quark and the charged (s)lepton sectors are complex at the GUT scale.
It is shown that, in this set up, up quark CEDM can be suppressed within the range where the current and future experiments have their sensitivity.
We briefly illustrate the simple realization of these particular forms of the modified sfermion universality with real up-(s)quark sector by  spontaneous CP violation in E$_6$ SUSY GUT with SU(2) flavor symmetry.

\end{abstract}

\end{titlepage}


\section{Introduction}

%
%
%

The low-energy supersymmetry (SUSY) is one of the leading candidates for the physics beyond the standard model (SM).
%
%
%
%
%
%
Since we have not seen yet any superpartner of the SM particle, SUSY must be a softly broken symmetry.
However once arbitrary soft SUSY breaking parameters 
are introduced into the model, SUSY contributions to several flavor changing neutral current (FCNC) processes and electric dipole moments (EDM) spoil the consistency between the observations and the SM predictions based on the Cabibbo-Kobayashi-Maskawa (CKM) matrix \cite{Cabibbo:1963yz} (so-called ``SUSY flavor problem" and ``SUSY CP problem" \cite{Gabbiani:1996hi}).

One of the solutions to avoid these problems is to assume certain mechanism which realizes that at some high energy scale sfermion masses are diagonal and universal in flavor space, and each scalar trilinear coupling is proportional to the corresponding Yukawa coupling (``universal solution") \cite{Kaplunovsky:1993rd, Dine:1981gu}.
However in order to avoid EDM constraints, an additional assumption is required.
We usually assume that all the SUSY breaking parameters and SUSY Higgs mass parameter $\mu $ are real.
One of the biggest mysteries is how to realize this assumption on CP invariance
for the SUSY breaking sector while the Yukawa couplings must be complex to obtain
the large Kobayashi-Maskawa (KM) phase. 

Another solution for these problems is to raise SUSY breaking scale \cite{Wells:2003tf}.
Then the SUSY contributions to the FCNC processes and EDMs can be decoupled (``decoupling solution").
However 
gluino and stop radiative corrections to the Higgs mass squared destabilize the weak scale.
It is possible to place stop masses in the weak scale in view of the naturalness \cite{Dimopoulos:1995mi}.
Unfortunately it has been pointed out that the large mass splitting between the 1st-2nd and the 3rd generation sfermions tends to make the 3rd generation sfermions masses squared negative through the two-loop renormalization group (RG)
 effects \cite{ArkaniHamed:1997ab}.

In this paper we examine another solution which has advantages of  both solutions noted above.
We place the 3rd generation SU(5) ten-plet sfermion masses in the weak scale in view of the naturalness.
The other sfermion masses are assumed to be universal and a TeV scale in order to weaken the constraints from the FCNC processes and EDMs.
To be concrete, we assume SU(5) ${\bold{ 10}}$-plet sfermion mass matrix $m_{\bold {10}}^2$ and ${\bar {\bold{ 5 }}}_{}$-plet sfermion mass matrix $m_{\bar {\bold {5}}}^2$ as
\begin{equation}
m_{\bold {10}}^2 =
  \left(
    \begin{array}{ccc}
       m_0^2 & 0 & 0 \cr
       0 & m_0^2 & 0 \cr
       0 & 0 & m_3^2
    \end{array}
  \right), \quad
m_{\bar{ \bold {5}}}^2= 
  \left(
    \begin{array}{ccc}
       m_0^2 & 0 & 0 \cr
       0 & m_0^2 & 0 \cr
       0 & 0 & m_0^2
    \end{array}
  \right). 
\label{eq:mod-uni}
\end{equation} 
We call this assumption as ``$\it{modified}$ sfermion universality"\cite{Maekawa:2002eh,Kim:2006ab,Kim:2008yta}.
In this set up if the gluino mass ($M_3$), the up-type Higgs mass ($m_{H_u}$) and $\mu $ as well as $m_3$ are the weak scale, the weak scale can be stabilized.%
\footnote{
In order to satisfy the lightest CP-even Higgs boson mass bound set by LEPI$\!$I one may employ $\it{maximal\, mixing}$ of stop sector \cite{Carena:1999xa}.
Another interesting option is $\it{light\, Higgs\, scenario}$ or $\it{inverted\, hierarchy}$ where one assumes that the heaviest CP-even Higgs boson resembles the SM Higgs boson \cite{Kane:2004tk}.
}
%
Since we can take $m_0 \gg m_3$ without destalizing the weak scale,\footnote{
Since SUSY contributions to the muon $g-2$ are decoupled, the anomaly \cite{Hagiwara:2006jt} is not explained if $m_0 \gtrsim $1TeV.
}
the constraints for the deviations from the ``modified universal solution" can be less severe than those for the deviations from the ``universal solution".

%
%

\begin{figure}[t]
 \begin{minipage}{0.5\hsize}
  \begin{center}
   \includegraphics[width=67mm]{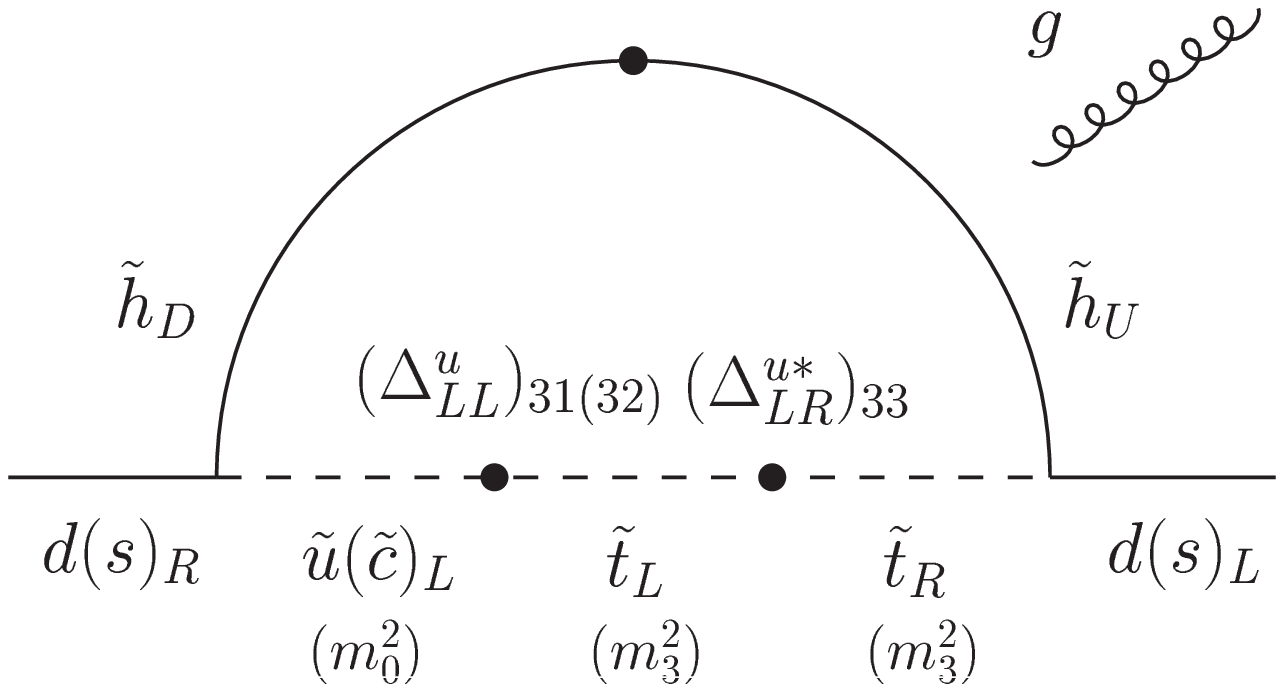} \\ (a)
  \end{center}
 \end{minipage}
 \begin{minipage}{0.5\hsize}
  \begin{center}
   \includegraphics[width=67mm]{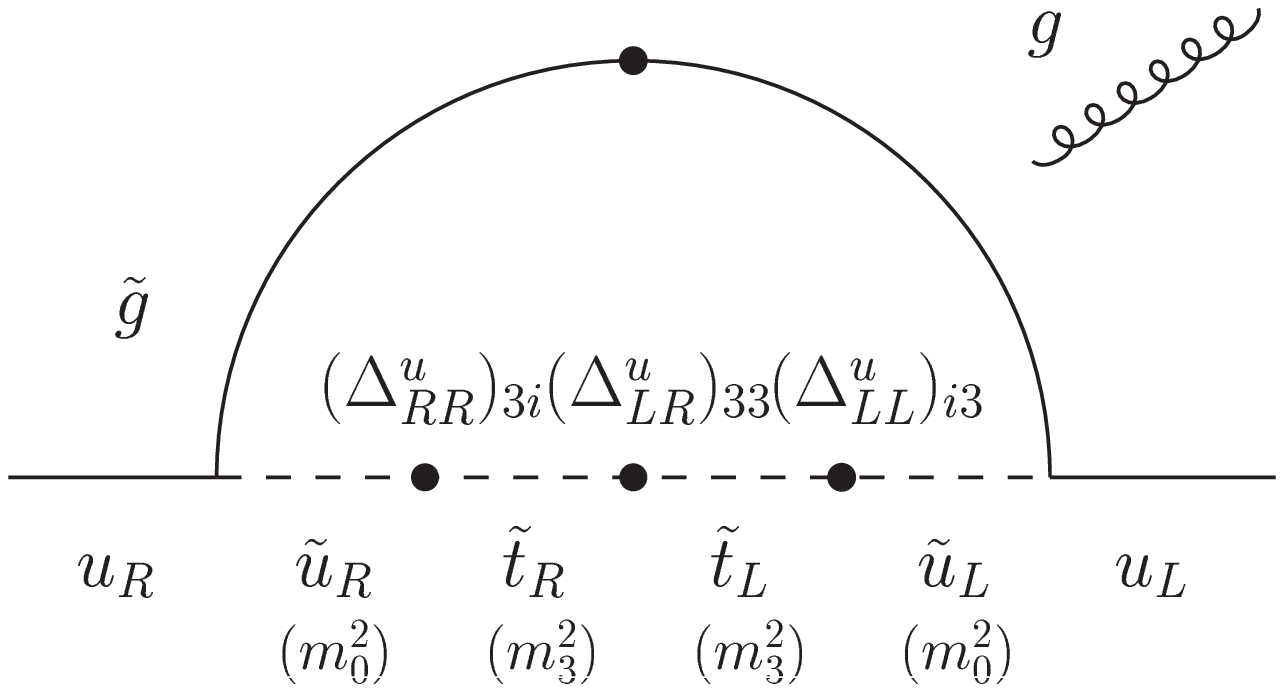} \\ (b)
  \end{center}
 \end{minipage}
  \caption{Non-decoupling SUSY contributions to CEDMs}
  \label{fig:non-decoupling SUSY contributions to cedms}
\end{figure}

Unfortunately, there are some SUSY contributions to the chromo-EDM (CEDM) of quarks which are not decoupled in the limit $m_0 \gg m_3$ as in Fig.\ref{fig:non-decoupling SUSY contributions to cedms}%
%
\footnote{
Throughout this paper we set QCD $\bar \theta$ parameter zero.
%
%
}.
The off-diagonal entries of the up-type squark soft mass matrix $(\Delta^u_{LL(RR)})_{i\not =j}$ in the so-called Super-CKM (SCKM) basis arise due to the $\it{modified}$ universal form of squark soft masses in the gauge eigen basis as%
\footnote{
Same for $\Delta^d_{LL}$.
}
\begin{equation}
(\Delta^u_{LL(RR)})_{i\not=j}
= [
  V_f^\dagger \left(\begin{array}{ccc}
  m_0^2 & 0 & 0 \cr
  0 & m_0^2 & 0 \cr
  0 & 0 & m_3^2
  \end{array}\right) V_f
  ]_{i\not=j}
=
\Delta m^2(V_f^*)_{3i} (V_f)_{3j}
\textnormal{\quad \footnotesize{$(f=U_L, U_R)$}}.
\label{eq:delta 13 in the modified universality}
\end{equation}
Here $\Delta m^2$=$ m_3^2- m_0^2$ and $V_f \textnormal{ \footnotesize{$(f=U_L, U_R)$}}$ are unitary matrices which diagonalize the up-type quark Yukawa coupling $Y_U$ of $Q Y_U U_R^C H_U$ as $V_{U_L}^T Y_U V_{U_R}^*$=$\hat Y_U$.
%
%
Since the mass insertion parameters $\delta^{u}_{LL(RR)}\equiv \Delta^{u}_{LL(RR)}/m_0^2$ do not vanish in the limit $m_0 \rightarrow \infty$, the SUSY contributions of Fig.\ref{fig:non-decoupling SUSY contributions to cedms} are not decoupled.
If we assume the hierarchy of $V_f$ as that of the CKM matrix, the mass insertion parameters $(\delta_{LL}^u)_{13}$ and  $(\delta_{RR}^u)_{31}$  become the order of $\lambda^3 \sim 0.01 $. (From here we set $\lambda = 0.22 $.)

Aside from the Fig.\ref{fig:non-decoupling SUSY contributions to cedms} (a) which entails the external quark Yukawa suppression%
\footnote{
With the parameters set used in the following discussions, the contribution to $d_d/e$ and $d_d^c$ from Fig. 1 (a) diagram are ${\cal O}(10^{-28})$cm that are an order of magnitude smaller than the current experimental bound.
}, Fig.\ref{fig:non-decoupling SUSY contributions to cedms} (b) prominently contributes to the up quark CEDM ($d_u^c$).
As an illustration, if we take $m_3 = M_3 = A_t \ll m_0 $ = 1TeV at the low-energy scale, the imaginary part of the product of $(\delta_{LL}^u)_{13}$ and  $(\delta_{RR}^u)_{31} $ is bounded by $^{199}$Hg (neutron) EDM as%
\footnote{
We assume $\text{Im}[M_3]=\text{Im}[A_t]=0$ for simplicity (it will be justified in the following discussions of the spontaneous CP violation).
The current experimental upper bounds for $^{199}$Hg and neutron EDMs are $|d_\text{Hg}| < 3.1 \times 10^{-29} \text{ e cm} $ \cite{Griffith:2009zz} and $|d_\text{n}| < 2.9 \times 10^{-26} \text{ e cm} $ \cite{Baker:2006ts}, respectively.
During the course of deriving Eq.\eqref{eq:hisano-shimizu}, we used $|d_u^c| < 6 \times 10^{-27} \text{ e cm} $ \cite{Griffith:2009zz} for $^{199}$Hg and $|d_u^c| < 1.8 \times 10^{-26} \text{ e cm} $ \cite{Hisano:2004tf} for neutron, as an example.
Other techniques and their constraints can be found in \cite{Latha:2009nq}, \cite{Ellis:2008zy} and references there in.
We do not assume any non-trivial cancellations among constituent particle (C)EDMs. 
Here we neglect a loop-originated factor that is a function of gaugino and sfermion mass ratio. E.g., in case of $m_3$=1TeV, the r.h.s. of Eq.\eqref{eq:hisano-shimizu} receives additional factor $\sim$ 3.
}
\begin{eqnarray}
\text{Im}[(\delta_{LL}^u)_{13}(\delta_{RR}^u)_{31}] \lesssim 3\times 10^{-7} \, (9 \times 10^{-7}) \cdot (\frac{m_3}{500\rm{GeV}})^2.
\label{eq:hisano-shimizu}
\end{eqnarray}
To obtain the KM phase, complex Yukawa couplings are assumed.
Then $(\delta_{LL}^u)_{13}$ and $(\delta_{RR}^u)_{31} $ derived from Eq.\eqref{eq:delta 13 in the modified universality}
can have different $\cal O$(1) phases generically and it is obviously incompatible with Eq.\eqref{eq:hisano-shimizu}.

%


One of the solutions to solve the SUSY CP problem is to introduce spontaneous CP violation in the flavor breaking sector.
%
%
More precisely we postulate that CP violation is related to the origin of Yukawa structure but not to SUSY breaking.
Then, first of all, CP phase will not primarily enter in the parameters that are independent of the flavor structure.
These are the gaugino masses, b- and $\mu$-parameters.
In contrast to them, Yukawa couplings, off-diagonal sfermion masses and A-terms can be expected to have CP phases.
%
Therefore to solve the SUSY CP problem in the modified sfermion masses, we need an additional assumption.
%
%
For example one of the interesting possibility on the flavor structure is Hermitian form \cite{Abel:2000hn}.
In this case $(A_u)_{33} $ is real in any base and phases among the product of $(\delta_{LL}^u)_{13}$ and $(\delta_{RR}^u)_{31}$
cancel with each other in SCKM base in good accuracy by virtue
of Hermitian nature.

In this paper, we propose an alternative solution for the SUSY CP problem in $\it{modified}$ sfermion masses.
%
We study a situation where the CP symmetry is effectively broken in the down-(s)quark (and the charged (s)lepton) sector but not in the up-(s)quark sector at the scale of spontaneous CP violation.
Then the above mentioned up quark CEDM induced via the stop loop can be well suppressed since parameters appearing in Fig.\ref{fig:non-decoupling SUSY contributions to cedms} (b) turn out to be almost real.
At the same time the KM phase is provided from the down-quark Yukawa
couplings which are complex.
%
Even if real up-(s)quark sector is assumed at the GUT scale it becomes complex at the low-energy.
We estimate this effects and show that the up quark CEDM induced via the stop loop can be suppressed within the range where the current and future experiment have their sensitivity.
Moreover we show that such an assumption can be simply realized by the spontaneous CP violation in E$_{\text 6}$ SUSY GUT with SU(2) flavor symmetry.

%
%
%

This paper is organized as follows.
In the next section we quantitatively examine our assumption for the SUSY CP problem in the $\it{modified}$ sfermion universality and estimate that EDM prediction.
%
%
%
In section 3, we give some comments for the simple realization of these particular forms of the $\it{modified}$ sfermion universality with real up-(s)quark sector by the spontaneous CP violation in the E$_6$ SUSY GUT with SU(2) flavor symmetry.
%
%
The last section is devoted to summary and discussion.

%

%
%
%

\section{CEDM bounds and real up-(s)quark sector}
\label{sec:SUSY CP problem}

As mentioned in the Introduction, the $\it{modified}$ sfermion universality is efficient to evade the most part of SUSY flavor problem and SUSY CP problem without destabilizing the weak scale.
At the same time, however, there is inevitable non-decoupling SUSY contribution to the up quark CEDM induced via the stop loop if their masses are placed in the weak scale in view of the naturalness.
%
%
To solve this difficulty, in this section, we quantitatively examine how taking the real up-(s)quark sector works to reduce the up quark CEDM induced via the stop loop.

%

%

\subsection{Real up-(s)quark sector}
\label{subsec:flavor dependent cp violation}

In this subsection we clarify our assumption of the real up-(s)quark sector.

Suppose the spontaneous CP violation in the flavor breaking sector.
Then the unified gaugino mass $M_{1/2}$, $\mu$ and  $b$, that are independent of the flavor symmetry breaking can be expected to be real and Yukawa couplings and A-terms become complex generically.
%
%
%
However in order to suppress the non-decoupling SUSY contribution to the up quark CEDM described in the previous section, here we assume that the up-type quark Yukawa couplings and corresponding A-term are real when these are induced by the flavor symmetry breaking whose scale is assumed to be near the GUT scale.
In summary we categorize Yukawa couplings, SUSY breaking terms and $\mu$-term into the following two sets at the GUT scale.

\begin{itemize}
\item Real 
\begin{itemize}
\item $M_{1/2}$, 
 $\mu$, $b$
\item ${\bold Y}_U$, ${\bold A}_U$, 
${\bold m}_{\tilde f}^2
\quad \textnormal{\scriptsize{$(f=Q, U_R)$}}$
\end{itemize}
\item Complex 
\begin{itemize}
\item 
${\bold Y}_f \, \textnormal{\scriptsize{$(f=D, \, E)$}}$,
${\bold A}_f \, \textnormal{\scriptsize{$(f=D, \, E)$}}$,
${\bold m}_{\tilde f}^2
\quad \textnormal{\scriptsize{$(f=D_R, \, L, \, E_R)$}}$
\end{itemize}
\label{test}
\end{itemize}
For A-terms, we assume that they duplicate the hierarchical structures of the corresponding Yukawa couplings but we do not require the exact proportionality among them, i.e., $Y_f$ and $A_f$ cannot be simultaneously diagonalized.
We introduce $A_0$ to denote the typical scale of $A$-terms as $(A_f)_{ij} \sim$ $(Y_f)_{ij} A_0$.
In connection with this assumption for A-terms, since $A_D$ is complex, we will take constraints from the down and strange quark CEDMs into account in the following analysis.
Even though these CEDM have decoupling features in the limit $m_0 \rightarrow \infty $, the recent update on the $^{199}$Hg EDM constrains relevant SUSY breaking parameters considerably.
%

\subsection{Quantitative evaluations}
\label{subsec:quantitative evaluations}

In this subsection, we quantitatively examine real up-(s)quark sector.

First of all, the up quark CEDM would be vanishing if the up sector has had only real parameters. 
However, even if ${\bold m}_{\tilde Q}^2$ is real at the input scale, it receives imaginary phases through the RG effects. 
Here we evaluate this effect schematically.
%
In the basis where $Y_U$ is diagonal, $Y_D$ can be written as $Y_D$=$PV_{\rm{CKM}}^* \hat Y_D$, where $\hat Y_D$ is diagonal Yukawa matrix, $P$=diag$(e^{i\phi_1},e^{i\phi_2},1)$ and $V_{\rm{CKM}}$ is the CKM matrix.
Then the CKM matrix emerges in the $\beta$-functions of $ {\bold m}_{\tilde Q}^2$ and the dominant contribution to $\text{Im} [ ({\bold m}_{\tilde Q}^2)_{i\not =j} ] $ is naively estimated as,%
%
%
\begin{eqnarray}
\text{Im}[ ({\bold m}_{\tilde Q}^2)_{i\not =j} ]
 \simeq
 \frac{1}{16 \pi^2} \textnormal{ln}(\frac{M_{GUT}}{M_{SUSY}}) {\rm Im[} (P^* V_{\rm{CKM}} \hat Y_D^2 V_{\rm{CKM}}^\dagger P m_{\bold{10}}^2 + {\it h.c.})_{ij}] \,.
\label{eq:rg effect on mq}
\end{eqnarray}
Therefore the Im[$(\delta^u_{LL})_{13} (\delta^u_{RR})_{31}$] turns out to be non vanishing.%
\footnote{
On the contrary to the ${\bold m}_{\tilde Q}^2$, the imaginary phases that arise from the RG effects are negligibly small for $ {\bold m}_{U_R^c}^2$ since they arise effectively at 2-loop level.
}
%
%
If we take $M_{\rm{GUT}}$=$2 \times 10^{16}$GeV, 
$M_{\rm{SUSY}}$=1TeV and $(\hat Y_D)_{33}$=0.13 (for $\tan \beta$=10 \cite{Xing:2007fb}), Eq.\eqref{eq:rg effect on mq} leads to Im[$(\delta^u_{LL})_{13}$\linebreak $ (\delta^u_{RR})_{31}$]$\simeq 1 \times10^{-7}$.
This value is a few times smaller than the bound of in Eq.\eqref{eq:hisano-shimizu}.
We note that this effect has an additional factor
$ (\tan\beta/10)^2$ and is non vanishing for large $m_0$.

For quantitative estimations we performed numerical analysis where the soft terms are pulled down to the low-energy scale using 2-loop RGEs \cite{Jack:1994rk} with the low-energy input \cite{Xing:2007fb}.
In the analysis, we specified following hierarchies for $Y_U$, $A_U$, $Y_D$, and $A_D$ at the GUT scale, motivated by the discussion given in the next section.
\begin{eqnarray}
Y_U \sim A_U \sim \left( 
\begin{array}{c c c}
\lambda ^6 & \lambda ^5 & \lambda ^3 \\
\lambda ^5 & \lambda ^4 & \lambda ^2 \\
\lambda ^3 & \lambda ^2 & 1 
\end{array}
\right)
\quad
Y_D \sim A_D \sim \left( 
\begin{array}{c c c}
\lambda ^6 & \lambda ^{5.5} & \lambda ^5 \\
\lambda ^5 & \lambda ^{4.5} & \lambda ^4 \\
\lambda ^3 & \lambda ^{2.5} & \lambda ^2 
\end{array}
\right)
\label{eq:Yu and Yd}
\end{eqnarray}
Also, we put real O(1) coefficient for each element of $Y_U$.
The coefficients are generated randomly within the interval 0.5 to 1.5 with + or - signs.
$Y_U$ generated in this way is used to determine the matrix
$V_f$ of Eq.\eqref{eq:delta 13 in the modified universality}.
We solved RGEs in the basis where $Y_U$ is diagonal.
For A-terms, we assigned another real O(1) coefficients randomly for $A_U$ as in the case of $Y_U$.
For $A_D$, we randomly put complex O(1) coefficients whose absolute values are constrained in the range from 0.5 to 1.5.

The up, down and strange quark CEDMs are evaluated using 1-loop formula,
\begin{eqnarray}
d^C_u &=& c \frac{\alpha_s}{4 \pi} \sum_{k=1}^6 \frac{M_3}{(\hat m_{\tilde u}^2)_{kk}}
 \{ 
(- \frac{1}{3} F_1 (x_k^u) - 3 F_2 (x_k^u) )
\text{Im} [(U_{\tilde u}^\dagger)_{1k} (U_{\tilde u})_{k4}]
 \} \\
d^C_d &=& c \frac{\alpha_s}{4 \pi} \sum_{k=1}^6 \frac{M_3}{(\hat m_{\tilde d}^2)_{kk}}
 \{ 
(- \frac{1}{3} F_1 (x_k^d) - 3 F_2 (x_k^d) )
\text{Im} [(U_{\tilde d}^\dagger)_{1k} (U_{\tilde d})_{k4}]
 \} \\
d^C_s &=& c \frac{\alpha_s}{4 \pi} \sum_{k=1}^6 \frac{M_3}{(\hat m_{\tilde d}^2)_{kk}}
 \{ 
(- \frac{1}{3} F_1 (x_k^d) - 3 F_2 (x_k^d) )
\text{Im} [(U_{\tilde d}^\dagger)_{2k} (U_{\tilde d})_{k5}]
 \} 
\label{eq:}
\end{eqnarray}
with fully diagonalizing up and down type squark mass matrices $m_{\tilde f}^2 $ ($f=u,d$) as $U_{\tilde f} m_{\tilde f}^2 U_{\tilde f}^\dagger $=$\hat m_{\tilde f}^2 $.
Here $c \sim $0.9 is the QCD correction, $F_3(x)$=$(x^2 -4x + 3 + 2 \text{Log}[x])/2 (1-x)^3$, $F_4(x)$=$(x^2-1-2 x \text{Log}[x])/2 (1-x)^3$ and $ x_k^f = \frac{M_3^2}{(\hat m_{\tilde f}^2)_{kk}} $.

\begin{figure}
\begin{center}
\includegraphics[width=90mm]{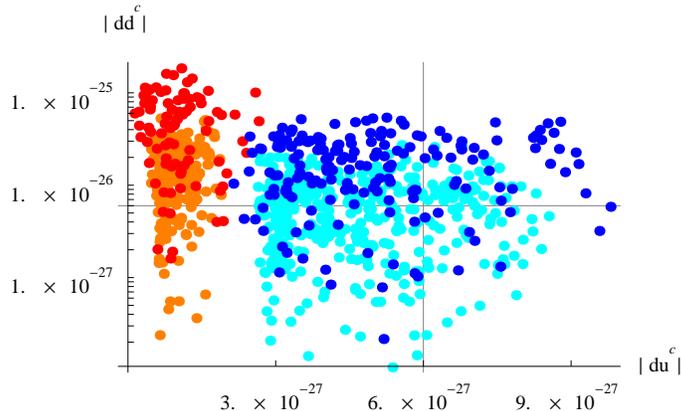}
\end{center}
\caption{up and down quark CEDMs}
\label{fig:edm results}
\end{figure}

The results are shown in Fig.\ref{fig:edm results}.
In the figure, the horizontal axis is absolute value of up quark CEDM and the vertical axis is absolute values of down quark CEDM.
Here, as an illustration, we took following 4 types of SUSY breaking parameters at the GUT scale;
[red : $m_0$=1TeV, $A_0$=-100GeV],
[orange : $m_0$=1TeV, $A_0$= -25GeV], 
[blue : $m_0$=2TeV, $A_0$=-100GeV], 
[cyan : $m_0$=2TeV, $A_0$=-25GeV]%
\footnote{
Here we made minus sign for $A_0$ only relevant to $A_U$(3,3).
}.
For the other relevant parameters, we set $m_3$=500GeV, $M_{1/2}$=200GeV, $\mu$=300GeV and $\tan \beta$=10.
In the figure, two straight lines represent current experimental upper bound ($|d_{u,d}^C|< 6 \times 10^{-27}$e cm \cite{Griffith:2009zz}) set by $^{199}$Hg EDM%
\footnote{
If we employ results of Refs.\cite{Hisano:2004tf} or \cite{Latha:2009nq}, $^{199}$Hg EDM constrains $|d_{u,d}^C|<3.5 \times 10^{-27}$ e cm and $|d_{u,d}^C|<3.2 \times 10^{-27}$ e cm, respectively.
}.
Therefore, the lower left area is allowed region.
In the plot, the points which induce strange quark CEDM that is larger than its experimental upper bound are omitted.
Here we required $|d_s^C| < 1.1 \times 10^{-25} {\rm e\, cm}$ \cite{Hisano:2004tf}.

From the figure, it is shown that the assumption of the real up-(s)quark sector has a chance to meet the experimental constraints, though it may requires some amount of tuning among O(1) parameters when A-term is not proportional to Yukawa coupling.
Regarding this point, we add some comments.
For $m_0$=2TeV cases (blue and cyan), we can see decoupling effects for $d_d^C$, that allows more points to enter the lower region.
On the other hand, since larger $m_0$ increases $\delta_{LL(RR)}^u$ for a while and decreases stop mass, $d_u^C$ increases when $m_0$ increases.
However, if we simultaneously increases $m_3$ some amount, more points will enter the allowed region in case of $m_0$=2TeV.
Also, since $^{199}$Hg EDM constrains difference of $d_u^C$ and $d_d^C$, some amount of accidental cancellation between them will allow more points to survive practically.
Finally, future deuteron EDM experiment whose proposed sensitivity $e |d_D| = (1\text{-}3) \times 10^{-27} \text{e cm}$ \cite{Semertzidis:2003iq} will constrain quark CEDMs at the level of $10^{-28}$ e cm.
Therefore we expect that the quark CEDMs that is discussed in this paper can be detected in future experiment, even if they meet current experimental bounds.

\section{Realization of modified universality and real up-(s)quark sector}
\label{sec:model of flavor dependent cp violation}

In this section we briefly illustrate the simple realization of the $\it{modified}$ sfermion universality with real up-(s)quark sector within the context of E$_6$ SUSY GUT with SU(2) flavor symmetry%
\footnote{
More detailed discussions and its interesting by-products will be given in the upcoming paper \cite{Ishiduki:2009}.
}.

\subsection{E$_6$ GUT}
\label{subsec:e6 gut}

We discuss the model within the frame work of E$_{\text 6}$ SUSY GUT \cite{Gursey:1975ki}.
Here the twisting mechanism among SU(5) ${\bar {\bold{ 5 }}}$ fields plays important multiple roles \cite{Maekawa:2002eh, Bando:1999km, Bando:2001bj}.

$\textbf{27}$ is the fundamental representation of the group, and in terms of E$_{ 6}\supset $SO(10)$\times $U(1)$_{V'}$ (and [SO(10)$\supset$SU(5)$\times$U(1)$_{V}$]), it is decomposed as%
\footnote{
Here acutes are used to distinguish different $\bar {\textbf{5}}$($\textbf{1}$)s.
}
\begin{eqnarray}
\textbf{27} = \textbf{16}_1 [ \textbf{10}_{-1} + \bar {\textbf{5}}_3 + \textbf{1}_{-5}] 
+ \textbf{10}_{-2} [ \textbf{5}_2 + \bar{\textbf{5}}_{-2}^\prime ] + \textbf{1}_4 [ \textbf{1}_0^\prime]
 \, .
\label{eq:e6 decomposition}
\end{eqnarray}

As one can see, $\textbf{27}$ incorporates two $\bar {\textbf{5}} $(${\textbf{1}} $)s of SU(5) and this nature allows  to produce various hierarchical structures of the Yukawa couplings in the SM from a single hierarchical structure of the original Yukawa couplings \cite{Bando:1999km, Bando:2001bj}.

We introduce the following superpotential
\begin{eqnarray}
W_{\textnormal E_{ 6}} \supset
Y_{ij}^H \Psi_i^{\textbf{27}} \Psi_j^{\textbf{27}} H^{\textbf{27}} + Y_{ij}^C \Psi_i^{\textbf{27}} \Psi_j^{\textbf{27}} C^{\textbf{27}}
\label{eq:E_6 Yukawa}
\end{eqnarray}
and assume the original Yukawa hierarchy as
\begin{eqnarray}
Y_{ij}^{H,C} \sim \left( 
\begin{array}{c c c}
\lambda ^6 & \lambda ^5 & \lambda ^3 \\
\lambda ^5 & \lambda ^4 & \lambda ^2 \\
\lambda ^3 & \lambda ^2 & 1 
\end{array}
\right) .
\label{eq:original Yukawa}
\end{eqnarray}
Here $\Psi_i^{\textbf{27}} \, (i=1\sim3) $ are the matter fields, and $H ^{\textbf{27}} $ and $C ^{\textbf{27}}$ are Higgs fields that break E$_{ 6}$ into SO(10) and SO(10) into SU(5), respectively.
Moreover here we assume that the MSSM Higgs doublets are included in $H ^{\textbf{27}} $ or $C ^{\textbf{27}}$.
More precisely, we also introduce adjoint field  $A ^{\textbf{78}} $ to realize the doublet-triplet Higgs mass splitting employing the Dimopoulos-Wilczek (DW) mechanism \cite{Dimopoulos:1981xm}, \cite{Maekawa:2001uk, Maekawa:2002bk}.
%

It turns out that once $H$ and $C$ acquire VEVs in the components of $\textbf{1}_4 ( \textbf{1}^\prime_0 )$ and $\textbf{16}_1 (\textbf{1}_{-5})$ of Eq.\eqref{eq:e6 decomposition}, respectively, it induces super heavy mass matrix of rank 3 among $\bold{ 5}_i, \bar {\bold{ 5 }}^\prime_i$ and $\bar {\bold{ 5 }}_i$ through the Yukawa coupling of Eq.\eqref{eq:E_6 Yukawa}.
Therefore the three degrees of freedom among $\bar {\bold{ 5 }}^\prime_i$ and $\bar {\bold{ 5 }}_i$ disappear below the GUT scale.
Consequently the up-quark Yukawa coupling ($Y_U$) remains as the original form of Eq.\eqref{eq:original Yukawa} but the down-quark ($Y_D$) and the charged lepton Yukawa couplings ($Y_E$) differ from it.
Importantly, the three massless modes among $\bar {\bold{ 5 }}^\prime_i$ and $\bar {\bold{ 5 }}_i$ mainly originate from the first two generations of $\Psi_{i\,\,(i=1\sim2)}^{27} $ because the 3rd generation ${\bar {\bold{ 5 }}}$ fields from $\Psi_{3}^{27} $   have large Yukawa couplings, i.e., large masses.
%
%
This feature is also important to make sfermion universality for ${\bar {\bold{ 5 }}}_{}$ sector as discussed in the following subsection.
%
As an example, when $\langle C \rangle/ \langle H \rangle \sim \lambda^{0.5}$ it ends up with the following milder hierarchies for $Y_D$ and $Y_E$.
%
%
\begin{eqnarray}
Y_D \sim Y_E^T \sim \left( 
\begin{array}{c c c}
\lambda ^6 & \lambda ^{5.5} &\lambda ^5 \\
\lambda ^5 & \lambda ^{4.5} &\lambda ^4 \\
\lambda ^3 & \lambda ^{2.5} &\lambda ^2 
\end{array}
\right).
\label{eq:yd and ye}
\end{eqnarray}
%
%
Note that the form of Eq.\eqref{eq:yd and ye} is adequate to reproduce not only the mass spectra of down-quark and charged lepton but also the large mixing of the Maki-Nakagawa-Sakata Matrix \cite{Maki:1962mu}.
Also Eq.\eqref{eq:yd and ye} leads to $\tan\beta \sim $10 which we used in the previous section.

\subsection{SU(2) flavor symmetry and modified universality}

Here we mention the emergence of the sfermion soft masses that have the $\it{modified}$ universal form of Eq.\eqref{eq:mod-uni} as discussed in 
Ref.\cite{Maekawa:2002eh}.

We attribute the original hierarchy of Eq.\eqref{eq:original Yukawa} to flavor symmetry breaking.
In order to naturally incorporate $\cal O$(1) top Yukawa coupling, here, we employ SU(2)$_{\text F}$ flavor symmetry and assume that the first two generations of matter fields are doublet $\Psi^{\textbf{27}}_a$ whereas $\Psi_3^{\textbf{27}}$ and Higgs fields are singlets.
We assume that the soft terms are mediated to the visible sector above the scale where the E$_{\text 6}$ and SU(2)$_{\text F}$ symmetries are respected, such as in the gravity mediation.
Then $\Psi^{\textbf{27}}_a$ acquires the soft mass ($m_0$) that is different from the soft mass of $\Psi^{\textbf{27}}_3$ ($m_3$) in general.
It guarantees the sfermion soft mass degeneracy at the GUT scale except for $\textbf{10}_3$ as
\begin{equation}
m_{\textbf {10}}^2 =
  \left(
    \begin{array}{ccc}
       m_0^2 & 0 & 0 \cr
       0 & m_0^2 & 0 \cr
       0 & 0 & m_3^2
    \end{array}
  \right), \quad
m_{\bar{ \textbf {5}}}^2= 
  \left(
    \begin{array}{ccc}
       m_0^2 & 0 & 0 \cr
       0 & m_0^2 & 0 \cr
       0 & 0 & m_0^2
    \end{array}
  \right),
\label{eq:modunif-sfermion-input}
\end{equation} 
because all the ${\bar {\bold{ 5 }}}$ fields come from a single fields $\Psi^{\textbf{27}}_a$.
This form is nothing but that of Eq.\eqref{eq:mod-uni}.%
\footnote{
Possible D-term contributions in case of gauged flavor symmetry and of the $E_6$ gauge symmetry are assumed to be negligible in this paper, 
though the large D-terms can induce too large
FCNC processes.
}
 Of course, the breaking of ${\rm SU(2)}_{\rm F}$ induces deviations from this form 
of the sfermion mass matrices, which can have complex components. 
We will discuss this effect in the next subsection.

\subsection{Spontaneous CP violation}
\label{subsec:spontaneous CP violation}

%
In this subsection we specify the origin of CP violating phase and derive the quark and lepton Yukawa couplings.

We assume that the flavor symmetry is broken by the VEV of the flavon fields $F_a$ and $\bar F^a$ that are fundamental and anti-fundamental under the SU(2)$_{\text F}$, respectively.
Once these $F_a$ and $\bar F^a$ acquire VEVs the interactions among matter fields, Higgs fields and flavon fields turn out as the part of the Yukawa coupling of Eq.\eqref{eq:E_6 Yukawa}.
Above all, we assume that the VEVs of the flavon fields entail the imaginary phase as $\langle F \bar F \rangle \sim e^{i\rho} \lambda^4$ \cite{Ishiduki:2009}.
As the results the CP and flavor symmetries are broken at the same time.
Using the SU(2)$_{\text F}$ gauge symmetry and its $D$-flatness condition, 
without loss of generality, we take the following VEVs,
\begin{eqnarray}
\langle F_a \rangle
\sim 
\left( 
\begin{array}{c}
0 \\
 e^{i\rho }\lambda ^2  
\end{array}
\right),
\indent
\langle \bar F_a \rangle
\sim 
\left( 
\begin{array}{c}
0 \\
\lambda ^2  
\end{array}
\right)
\label{<F_a>},
\end{eqnarray}
where only $\langle F_2 \rangle $ acquires imaginary phase. (In this paper, we
take the cutoff $\Lambda$=1.)
In the following, we
assume that this phase is the only CP violating phase in this model.

If the  MSSM up-type Higgs $H_U$ comes from $H^\bold{27}$, then $Y_U$=$Y_H$.
In this case when an appropriate symmetry forbids $F$ to participate the generation of $Y^H$ and $m_{\bf 10}^2$ but allows $F$ to participate the generation of $Y^C$, then the real $m_{\bf 10}^2$, the real up-type quark Yukawa couplings
and therefore, the real up-type A-term can be obtained. 
Moreover, complex down-type quark Yukawa couplings, which are important to induce
KM phase,  can be obtained.
%
%
%
These structures are nothing but we would like to obtain to solve the SUSY CP problems in $\it{modified}$ sfermion universality.

%
%
%




We presented explicit discrete symmetry in Table \ref{tb:field contents and charge assignment} to satisfy these features
%
\footnote{
We implicitly assume additional ${\overline{\bf 27}}$ representation Higgses to satisfy the D-flatness conditions of $E_6$ and additional odd number SU(2) doublet(s) to avoid an SU(2) anomaly \cite{Witten:1982fp}.
}.
\begin{table}[]
\begin{center}
\begin{tabular}{cccccccccccccccc}
\toprule
 &$\Psi_a$&$\Psi_3$&$F_a$&$\bar F^a$&$H$&$C$&$A$\\
\midrule
E$_{\text 6}$ & $\bold{27}$ & $\bold{27}$ &$\bold{1}$ & $\bold{1}$& $\bold{27}$&$\bold{27}$ &$\bold{78}$  \\
SU(2)$_{\text 2}$& $\bold{2}$ & $\bold{1}$ & $\bold{2}$ &$\bar {\bold{2}}$&$\bold{1}$&$\bold{1}$&$\bold{1}$  \\
$Z_3$ & 0 & 0 & 1 & 0 & 0 & 2 & 0  \\
\bottomrule
\end{tabular}
\end{center}
\caption{Field contents and charge assignment}
\label{tb:field contents and charge assignment}
\end{table}
With this assignment, following interactions
\begin{eqnarray}
Y_H &:&
\left( 
\begin{array}{ccc}
0            &  \Psi_a A \Psi^a  & 0 \\
\Psi_a A \Psi^a & (\bar F^a \Psi_a)^2 & \bar F^a \Psi_a  \Psi^3 \\
0            & \Psi_3 \bar F^a \Psi_a & \Psi_3  \Psi_3
\end{array}
\right) H 
\label{eq:yhz6} \\
\nonumber \\
Y_C &:&
\left( 
\begin{array}{ccc}
0            & F^a \Psi_a \bar F^b \Psi_b & F^a \Psi_a \Psi_3 \\
\bar F^a \Psi_a F^b \Psi_b & 0 & 0 \\
\Psi_3 F^a \Psi_a & 0 & 0
\end{array}
\right) C
\label{eq:ycz6}
\end{eqnarray}
are responsible for the Yukawa couplings of $Y_H$ and $Y_C$ after the SU(2)$_{\text F}$ symmetry breaking%
\footnote{
We ignore the ${\cal O}(\lambda^{8})$ contribution $F\bar F (F^a\Psi_a)^2$ to (1,1) element of Eq.\eqref{eq:yhz6} for simplicity.
Though it acquires CP violating phase, its contributions to the up quark CEDM does not change the results presented in this paper.
%
}.


First of all, when the Higgs fields and the flavon fields acquire VEVs, these interactions induce following mass matrix for ${\bold 5}_{i}$, $\bar {\bold 5}_{i}'$ and $\bar {\bold 5}_{i}$,
\begin{eqnarray}
\bordermatrix{ 
   & \bar {\bold{ 5 }}^\prime_1 &\bar {\bold{ 5 }}^\prime_2 & \bar {\bold{ 5 }}^\prime_3 & {\bar {\bold{ 5 }}}_1 &{\bar {\bold{ 5 }}}_2 &{\bar {\bold{ 5 }}}_3\cr
\bold{ 5 }_1& 0            & \alpha d \lambda^5 & 0           & 0             & f e^{i\delta} \lambda^{5.5} & g e^{i\delta} \lambda^{3.5} \cr
\bold{ 5 }_2& -\alpha d \lambda^5 & c \lambda^4 & b \lambda^2 & f e^{i\delta} \lambda^{5.5} & 0  & 0 \cr
\bold{ 5 }_3& 0 & b \lambda^2 & a & g e^{i\delta} \lambda^{3.5} & 0 & 0 \cr}
\langle H \rangle.
\label{eq:z655bp5b}
\end{eqnarray}
Here we assume $\langle A \rangle \sim \lambda^5$ and $\langle C \rangle/\langle H \rangle \sim \lambda^{1.5}$.
Also we restore the $\cal O$(1) coefficient of the terms which appear  in Eq.\eqref{eq:yhz6} and Eq.\eqref{eq:ycz6}.
These coefficients, $a,b,c,d,f$ and $g$, are all real because of the CP symmetry.
It is important to note that $\alpha=1$ for $D_R^C$ component of $\bar {\bold{ 5 }}^\prime_{}$ and  $\alpha=0$ for $L$ component of $\bar {\bold{ 5 }}^\prime_{}$ since the (1,2) and (2,1) elements of Eq.\eqref{eq:z655bp5b}  originate from the B-L conserving VEV of $A$.
%
%
Then L[$\bar {\bold 5}'_1$] component becomes purely massless mode and L[$\bar {\bold 5}_3$] can not contain the massless mode whose main mode is L[$\bar {\bold 5}'_1$].
Therefore it is mandatory to manage the Higgs potential so that L[$C^{ {\bold 16(\bar 5)}} $] possesses a fraction of $H_D$ otherwise determinant of $Y_E^T$ vanishes.
In the following we assume that $H_D \sim {\text L}$[$H^{\bold {10} (\bar {\bold{5}}^\prime)}$]+$\lambda^{0.5}{\text L}$[$C^{{\bold{ 16} \bar {\bold{(5)}}}}$].
%

Now we derive the Yukawa couplings $Y_U$, $Y_D$ and $Y_E$ in a semi-analytical way assuming that the each $\cal O$(1) coefficient does not alter the hierarchical structures that originate from the VEVs of the Higgs and the flavon fields.
First of all $Y_U$ is simply derived from Eq.\eqref{eq:yhz6} extracting corresponding components as%
\footnote{
Now we take following conventions where $d_0$ denotes the original $\cal O$(1) coefficient that appears in $\Psi_a A \Psi^a H$;
$d_0 \langle A_{B-L} \rangle D_R^C [ {\bar {\bold{ 5 }}}^\prime]$ = $d \lambda^5 D_R^C [ {\bar {\bold{ 5 }}}']$,
$d_0 \langle A_{B-L} \rangle U_R^C [ { {\bold{ 10 }}}]$ = $\frac{-d}{2} \lambda^5 U_R^C [ { {\bold{ 10 }}}]$,
$d_0 \langle A_{B-L} \rangle D_R^C [ {\bar {\bold{ 5 }}}]$ = $\frac{-d}{2} \lambda^5 D_R^C [ {\bar {\bold{ 5 }}}]$, and 
$d_0 \langle A_{B-L} \rangle L[ {\bar {\bold{ 5 }}}]$ = $\frac{-3d}{2} \lambda^5 L [ {\bar {\bold{ 5 }}}]$.
}
\begin{eqnarray}
Y_U =
\bordermatrix
{ 
               & {U_{R}^C}_1      & {U_{R}^C}_2      & {U_{R}^C}_3 \cr
Q_{1} & 0 & -\frac{1}{2} d \lambda^{5} & 0 \cr
Q_{2} & \frac{1}{2} d \lambda^5 & c \lambda^{4} & b \lambda^{2} \cr
Q_{3} & 0 & b \lambda^{2} & a \cr
}.
\label{eq:z6Yu}
\end{eqnarray}
For $Y_D$ and $Y_E$, we first derive the relation between the gauge eigen modes of $\bar {\bold 5}_{i}$ and $\bar {\bold 5}_{i}'$ and mass eigen modes from Eq.\eqref{eq:z655bp5b}.
Then we replace $\bar {\bold 5}_{i}$ and $\bar {\bold 5}_{i}'$ that appear in Eq.\eqref{eq:yhz6} and Eq.\eqref{eq:ycz6} in terms of massless modes.
Also at the same step we replace L[$\bar {\bold 5}_{H}'$] and L[$\bar {\bold 5}_{C}$] as $H_D$ and $\beta \lambda^{0.5} H_D$ respectively.
Here $\beta $ denotes another $\cal O$(1) coefficient which enters in this step.
Correcting the leading order contributions we finally arrive the following Yukawa couplings
\begin{align}
&Y_D = \nonumber \\
&\bordermatrix
{ 
               & {D_{R}^C}_1      & {D_{R}^C}_2      & {D_{R}^C}_3 \cr
Q_{1} & -\{(\frac{bg-af}{g})^2 \frac{1}{ac-b^2} +1 \}\frac{g}{a} g \beta e^{2i\delta} \lambda^{6} & -\frac{bg-af}{g} \frac{d}{ac-b^2} g \beta e^{i\delta} \lambda^{5.5}& -\frac{1}{2} d \lambda^5 \cr
Q_{2} & (\frac{d}{2}- \frac{d}{ac-b^2} \frac{bg-af}{g} b) \lambda^5 & (-\frac{ad^2}{ac-b^2} \frac{b}{g} e^{-i\delta} + \beta f e^{i\delta}) \lambda^{4.5} & (\frac{ac-b^2}{a} + \frac{bg-af}{g} \frac{b}{a}) \lambda^{4} \cr
Q_{3} & -\frac{ad}{ac-b^2} \frac{bg-af}{g} \lambda^3 & (-\frac{ad^2}{ac-b^2} \frac{a}{g} e^{-i\delta} + \beta g e^{i\delta}) \lambda^{2.5} & \frac{bg-af}{g} \lambda^{2} \cr
},
\nonumber \\
\label{eq:z6Yd}
\end{align}
\begin{eqnarray}
Y_E^T =
\bordermatrix
{ 
               & L_1      & L_2      & L_3 \cr
{E_R^C}_{1} & -\{(\frac{bg-af}{g})^2 \frac{1}{ac-b^2} +1 \}\frac{g}{a} g \beta e^{2 i\delta} \lambda^{6} & 0 & -\frac{3}{2} d \lambda^5 \cr
{E_R^C}_{2} & \frac{3}{2} d \lambda^5 & \beta f e^{i\delta} \lambda^{4.5} & (\frac{ac-b^2}{a} + \frac{bg-af}{g} \frac{b}{a}) \lambda^{4} \cr
{E_R^C}_{3} & 0 & \beta g e^{i\delta} \lambda^{2.5} & \frac{bg-af}{g} \lambda^{2} \cr
}.
\nonumber \\
\label{eq:z6Ye}
\end{eqnarray}
%
%
As a whole, $Y_U$ is real whereas $Y_D$ and $Y_E$ are complex and possess non-removable phases as one can see in Eqs.\eqref{eq:z6Yu}, \eqref{eq:z6Yd} and \eqref{eq:z6Ye}.
More detailed discussions and its interesting phenomenological implications are presented in \cite{Ishiduki:2009}.

Finally, we comment on the parameters $m_{\bf 10}^2$. As noted before, the complex
non-vanishing VEV $\langle F\rangle$ may induce complex $m_{\bf 10}$ generically.
For example, the terms $\tilde m^2 \Psi_3^\dagger \epsilon^{ab} F_a \Psi_b$ and $\tilde m^2 \Psi_3^\dagger(F^\dagger)^a\Psi_a$ result
in complex $(m_{\bf 10}^2)_{13}$ and $(m_{\bf 10}^2)_{23}$, respectively, 
after developping the complex VEV of $F$.
However, in this explicit model, the $Z_3$ discrete symmetry forbids these terms,
so the complex components in $m_{\bf 10}^2$ become sufficiently small.
Since the FCNC constraints from the deviations by the non-vanishing VEVs of $F$ and 
$\bar F$ and by the $E_6$ breaking effects are argued in Ref. \cite{Maekawa:2002eh},
we do not repeat the argument here. But the FCNC constraints can be satisfied
in this model.

%

\section{Summary and discussion}
\label{sec:summary and discussion}

In this paper we discussed the SUSY CP problem that arises when the sfermion soft mass universality is $\it{modified}$ as in Eq.\eqref{eq:mod-uni} and $m_3$ is placed in the weak scale in view of the naturalness.

In order to suppress up quark CEDM induced by the weak scale stop pair we employed certain type of favor structure where the parameters of up-(s)quark sector are real whereas those of down-(s)quark and charged (s)lepton sectors are complex at the GUT scale.
%
%
It has shown that, in this set up, the up quark CEDM can be suppressed within the range where the current and future experiment have their sensitivity.
%
%

In Sec.\ref{sec:model of flavor dependent cp violation}, we briefly illustrated the simple realization of these particular forms of the  $\it{modified}$ sfermion universality with real up-(s)quark sector by the spontaneous CP violation in the E$_6$ SUSY GUT with SU(2) flavor symmetry.
By virtue of the twisting among the SU(5) $\bar {\bold{ 5 }}$ matter fields, those are easily obtained.
The model building of the spontaneous CP violation in the E$_6$ SUSY GUT is an interesting topic itself which is discussed in the upcoming paper \cite{Ishiduki:2009}.

\section*{Acknowledgment}

S.K. and K.S. are supported in part by Grants-in-Aid for JSPS fellows.
N.M. is supported in part by JSPS Grants-in-Aid for Scientific Research.
This research was partially supported by the Grant-in-Aid for Nagoya University Global COE Program, "Quest for Fundamental Principles in the Universe: from Particles to the Solar System and the Cosmos", from the Ministry of Education, Culture, Sports, Science and Technology of Japan.
The Feynman diagrams in the paper were made with \texttt{JaxoDraw} \cite{Binosi:2003yf}.

\end{document}